\documentclass[12pt]{article}

\usepackage{graphicx}
\usepackage{float}

\def\gtwid{\mathrel{\raise.3ex\hbox{$>$\kern-.75em\lower1ex\hbox{$\sim$}}}}
\def\ltwid{\mathrel{\raise.3ex\hbox{$<$\kern-.75em\lower1ex\hbox{$\sim$}}}}
\def\square{\kern1pt\vbox{\hrule height 1.2pt\hbox{\vrule width 1.2pt\hskip 3pt
   \vbox{\vskip 6pt}\hskip 3pt\vrule width 0.6pt}\hrule height 0.6pt}\kern1pt}

\begin{document}

\begin{titlepage}

\begin{flushright}
UFIFT-QG-18-02
\end{flushright}

\vskip 1cm

\begin{center}
{\bf The Graviton Tail almost Completely Wags the Dog}
\end{center}

\vskip 1cm

\begin{center}
S. P. Miao$^{1*}$, T. Prokopec$^{2\star}$ and R. P. Woodard$^{3\dagger}$
\end{center}

\vskip .5cm

\begin{center}
\it{$^{1}$ Department of Physics, National Cheng Kung University \\
No. 1, University Road, Tainan City 70101, TAIWAN}
\end{center}

\begin{center}
\it{$^{2}$ Institute for Theoretical Physics, Spinoza Institute \& EMME$\Phi$ \\
Utrecht University, Postbus 80.195, 3508 TD Utrecht, NETHERLANDS}
\end{center}

\begin{center}
\it{$^{3}$ Department of Physics, University of Florida,\\
Gainesville, FL 32611, UNITED STATES}
\end{center}

\vspace{1cm}

\begin{center}
ABSTRACT
\end{center}
One graviton loop corrections to the vacuum polarization on de
Sitter show two interesting infrared effects: a secular enhancement 
of the photon electric field strength and a long range running of 
the Coulomb potential. We show that the first effect derives solely 
from the ``tail'' term of the graviton propagator, but that the
second effect does not. Our result agrees with the earlier 
observation that the secular enhancement of massless fermion mode 
functions derives from solely from the tail term. We discuss the 
implications this has for the important project of generalizing to 
quantum gravity the Starobinsky technique for summing the series of 
leading infrared effects from inflationary quantum field theory.

\begin{flushleft}
\end{flushleft}

\vspace{1cm}

\begin{flushleft}
$^{*}$ e-mail: spmiao5@mail.ncku.edu.tw \\
$^{\star}$ email: T.Prokpec@uu.nl \\
$^{\dagger}$ e-mail: woodard@phys.ufl.edu
\end{flushleft}

\end{titlepage}

\section{Introduction}

It has long been clear that there is something peculiar 
about long wavelength gravitons on cosmological backgrounds
\cite{Lifshitz:1945du}. Unlike photons, which are precluded by 
conformal invariance from locally perceiving the expansion of the 
Universe, inflationary expansion leads to the production of gravitons 
\cite{Grishchuk:1974ny,Ford:1977dj}. This process is the source of 
the tensor power spectrum predicted by primordial inflation 
\cite{Starobinsky:1979ty}.

Long wavelength gravitons also make a peculiar contribution to the
retarded propagator, which DeWitt and Brehme famously denoted as the
``tail term'' \cite{DeWitt:1960fc}. Unlike the usual delta function
{\it on} the past light-cone, the tail contribution is nonzero {\it 
inside} the past light-cone \cite{Chu:2011ip}. This fact has great
relevance to computations of gravitational radiation reaction in
binary mergers \cite{Tanaka:1996ht,Mino:1996nk,Quinn:1996am}. It is
also responsible for the curious infrared ``running'' of the Newtonian 
potential induced by the one loop gravitational vacuum polarization of 
conformal matter on de Sitter background \cite{Wang:2015eaa,Frob:2016fcr},
\begin{equation}
\Psi = -\frac{GM}{a r} \Biggl\{ 1 + \frac{4 G}{15 \pi a^2 r^2} + 
\frac{2 G H^2}{5 \pi} \, \ln(a H r) + O(G^2)\Biggr\} \; . \label{confNewt}
\end{equation}
Here $H$ is the Hubble constant, $a = e^{H t}$ is the de Sitter scale 
factor and $r$ is the co-moving position. The fractional correction of
$\frac{4 G}{15 \pi a^2 r^2}$ is just the de Sitter descendant of the flat 
space effect which has long been known \cite{Radkowski:1970,Capper:1974ed}.
The new term proportional to $G H^2$ is specific to nonzero Hubble constant 
and causes perturbation theory to break down, both for large $r$ and at 
late times. Even though conformal matter induces almost the same vacuum 
polarization, in de Sitter conformal coordinates, as in flat space, the 
gravitational {\it response} to that source is very different on account 
of the strong de Sitter tail term.

Analytic continuation carries the tail term of the retarded propagator into
the tail part of the Feynman propagator which can mediate quantum graviton
effects to other particles \cite{Tsamis:1992xa,Woodard:2004ut}. An important
example is the one graviton contribution to the electromagnetic vacuum 
polarization \cite{Leonard:2013xsa}. This induces an infrared running of 
the Coulomb potential similar to (\ref{confNewt}) \cite{Glavan:2013jca},
\begin{equation}
\Phi = \frac{Q}{4\pi r} \Biggl\{ 1 + \frac{2 G}{3 \pi a^2 r^2} + 
\frac{2 G H^2}{\pi} \, \ln(a H r) + O(G^2)\Biggr\} \; . \label{gravCoul}
\end{equation}
As with the Newtonian potential (\ref{confNewt}), the fractional correction
$\frac{2 G}{3 \pi a^2 r^2}$ is just the de Sitter analogue of what happens
in flat space \cite{Leonard:2012fs}, while the new term proportional to 
$G H^2$ causes perturbation theory to break down at large $r$ and at late 
times. The gravitational vacuum polarization on de Sitter also causes a 
secular enhancement of the electric field of a plane wave photon 
\cite{Wang:2014tza},
\begin{equation}
F^{\rm 1~loop}_{0i} \longrightarrow \frac{2 G H^2}{\pi} \, \ln(a) \times
F^{\rm tree}_{0i} \; . \label{gravphot}
\end{equation}
Like (\ref{gravCoul}), this result signals a late time breakdown of 
perturbation theory.

A common feature in all three results (\ref{confNewt}), (\ref{gravCoul}) and
(\ref{gravphot}) is the breakdown of perturbation theory when $\ln(a) \sim 
\frac1{G H^2}$. Uncovering what happens after this time requires going beyond
perturbation theory. For the very similar infrared logarithms of scalar 
potential models Starobinsky has developed a stochastic formalism 
\cite{Starobinsky:1986fx} which exactly reproduces the leading infrared 
logarithms at each loop order \cite{Woodard:2005cw,Tsamis:2005hd}, and can
be summed to elucidate the nonperturbative regime \cite{Starobinsky:1994bd}.
The same technique can be applied to a Yukawa-coupled scalar \cite{Miao:2006pn},
and to scalar quantum electrodynamics \cite{Prokopec:2007ak}. However, it has
not yet been generalized to quantum gravity.

The obstacle to applying Starobinsky's formalism has been the derivative 
interactions of quantum gravity. These frustrate the proof \cite{Woodard:2005cw,
Tsamis:2005hd} that works for scalar potential models. Derivative interactions
also mean that the lowest order renormalization counterterms contribute at
leading logarithm order, which means that dimensional regularization must be 
retained until a fully renormalized result is obtained \cite{Miao:2008sp}.
The problem remains, despite notable progress understanding the simpler 
derivative interactions of nonlinear sigma models \cite{Kitamoto:2010et,
Kitamoto:2011yx}.

A notable advance was the discovery \cite{Miao:2008sp} that only the tail
part of the graviton propagator is responsible for the secular enhancement
of massless fermions on de Sitter background \cite{Miao:2005am,Miao:2006gj}.
The purpose of this paper is to see if the tail term alone also explains the 
secular enhancement of dynamical photons (\ref{gravphot}) and the logarithmic 
running of the Coulomb potential (\ref{gravCoul}). In section 2 we review the 
relevant Feynman rules and identify precisely those parts of the vacuum 
polarization which are responsible for the two effects. Section 3 computes the 
tail contribution to the vacuum polarization. Our results are discussed in 
section 4.

\section{Notation}

The purpose of this section is to review notation. We begin with the Feynman
rules which were used to compute the vacuum polarization \cite{Leonard:2013xsa}.
This is where we define the ``tail'' part of the graviton propagator which plays
a central work in this study. We also describe how the tensor structure of the
vacuum polarization is represented using two structure functions, and we give
the order $G H^2$ contributions to these structure functions which are 
responsible for the enhancement of dynamical photons (\ref{gravphot}) and
the logarithmic running of the Coulomb potential (\ref{gravCoul}).

\subsection{Feynman Rules}

The Lagrangian relevant to our study is,
\begin{equation}
\mathcal{L} = \frac{[R \!-\! (D\!-\!2)(D\!-\!1) H^2] \sqrt{-g}}{16 \pi G} 
-\frac14 F_{\mu\nu} F_{\rho\sigma} g^{\mu\rho} g^{\nu\sigma} \sqrt{-g} + 
\Delta \mathcal{L} + \mathcal{L}_{GF} \; . \label{Lag}
\end{equation}
Here $D$ is the spacetime dimension, $H$ is the de Sitter Hubble constant and 
$G$ is Newton's constant. The two counterterms we require are,
\begin{equation}
\Delta \mathcal{L} = \overline{C} H^2 F_{\mu\nu} F_{\rho\sigma} g^{\mu\rho}
g^{\nu\sigma} \sqrt{-g} + \Delta C H^2 F_{ij} F_{k\ell} g^{ik} g^{j\ell} 
\sqrt{-g} \; . \label{DLag}
\end{equation}
The noninvariant term (Roman indices are purely spatial) proportional to 
$\Delta C$ is required because of de Sitter breaking in the graviton sector 
\cite{Leonard:2013xsa,Glavan:2015ura}. Our electromagnetic and gravitational 
gauge fixing terms are \cite{Tsamis:1992xa,Woodard:2004ut},
\begin{equation}
\mathcal{L}_{GF} = -\frac12 a^{D-4} \Bigl[ \eta^{\mu\nu} A_{\mu , \nu} \!-\! 
(D \!-\!4) H a A_0\Bigr]^2 -\frac12 a^{D-2} \eta^{\mu\nu} F_{\mu} F_{\nu} \; ,
\label{LGF}
\end{equation}
where $a \equiv -\frac1{H \eta}$ is the de Sitter scale factor (at conformal
time $\eta$) and the gravitational term is,
\begin{equation}
F_{\mu} \equiv \eta^{\rho\sigma} \Bigl[h_{\mu\rho ,\sigma} \!-\! \frac12 
h_{\rho\sigma , \mu} \!+\! (D \!-\! 2) H a h_{\mu\rho} \delta^0_{~\sigma} \Bigr]
\; . \label{Fmu}
\end{equation}
Here and henceforth $h_{\mu\nu}$ is the conformally transformed graviton field
whose indices are raised and lowered with the (spacelike) Minkowski metric,
\begin{equation}
g_{\mu\nu} \equiv a^2 \widetilde{g}_{\mu\nu} \equiv a^2 \Bigl[ \eta_{\mu\nu} + 
\kappa h_{\mu\nu} \Bigr] \qquad , \qquad \kappa^2 \equiv 16 \pi G \; .
\label{graviton}
\end{equation}

Our gauge breaks de Sitter invariance but it does provide the simplest 
possible expressions for the photon and graviton propagators. They each take 
the form of a sum of constant tensor factors times scalar propagators,
\begin{eqnarray}
i\Bigl[\mbox{}_{\mu} \Delta_{\rho}\Bigr](x;x') & = &
\overline{\eta}_{\mu\rho} \!\times\! a a' i\Delta_B(x;x') - \delta^0_{~\mu} 
\delta^0_{~\rho} \!\times\! a a' i \Delta_C(x;x') \; , \qquad \label{photprop} \\
i\Bigl[\mbox{}_{\mu\nu} \Delta_{\rho\sigma}\Bigr](x;x') & = & 
\sum_{I=A,B,C} \Bigl[\mbox{}_{\mu\nu} T^I_{\rho\sigma}\Bigr] \!\times\! 
i\Delta_I(x;x') \; , \label{gravprop}
\end{eqnarray}
where $\overline{\eta}_{\mu\nu} \equiv \eta_{\mu\nu} + \delta^0_{~\mu}
\delta^0_{~\nu}$ is the spatial part of the Minkowski metric. The gravitational
tensor factors are,
\begin{eqnarray}
\Bigl[\mbox{}_{\mu\nu} T^A_{\rho\sigma}\Bigr] & = & 
2 \overline{\eta}_{\mu (\rho} \overline{\eta}_{\sigma) \nu} - \frac{2}{D \!-\! 3} \, 
\overline{\eta}_{\mu\nu} \overline{\eta}_{\rho\sigma} \; , \label{TA} \\
\Bigl[\mbox{}_{\mu\nu} T^B_{\rho\sigma}\Bigr] & = & -4 \delta^0_{~(\mu}
\overline{\eta}_{\nu) (\rho} \delta^0_{~\sigma)} \; , \label{TB} \\ 
\Bigl[\mbox{}_{\mu\nu} T^C_{\rho\sigma}\Bigr] & = & \frac{2}{(D\!-\!2) 
(D \!-\! 3)} \Bigl[ (D \!-\! 3) \delta^0_{~\mu} \delta^0_{~\nu} \!+\! 
\overline{\eta}_{\mu\nu} \Bigr] \Bigl[ (D \!-\! 3) \delta^0_{~\rho} \delta^0_{~\sigma} 
\!+\! \overline{\eta}_{\rho\sigma} \Bigr] \; . \label{TC} \qquad
\end{eqnarray}
Here and henceforth parenthesized indices are symmetrized.

It is useful to expand the three scalar propagators in progressively less and less
singular terms,
\begin{equation}
i\Delta_I(x;x') = \frac{i\Delta(x;x')}{(a a')^{\frac{D}2-1}} + i\delta \Delta_I(x;x') 
+ i\Delta_{\Sigma I}(x;x') \quad , \quad I = A, B, C \; . \label{scalarprops}
\end{equation}
Here the massless scalar propagator in flat space is
\begin{equation}
i\Delta(x;x') = \frac{ \Gamma(\frac{D}2 \!-\! 1)}{4 \pi^{\frac{D}2} \Delta x^{D-2}}
\quad , \quad \Delta x^2(x;x') \equiv \Bigl\Vert \vec{x} \!-\! \vec{x}' \Bigr\Vert^2 - 
\Bigl(\vert \eta \!-\! \eta'\vert \!-\! i \epsilon\Bigr)^2 \; . \label{simpleprop}
\end{equation}
Note that $i\Delta(x;x')$ has the leading, $1/\Delta x^{D-2}$ singularity. The three
$1/\Delta x^{D-4}$ terms are,
\begin{eqnarray}
\lefteqn{(a a')^{\frac{D}2 - 2} i\delta \Delta_A(x;x') = \frac{H^2}{4 \pi^{\frac{D}2}} 
\Biggl\{ \frac{ \Gamma(\frac{D}2 \!+\! 1)}{2 (D \!-\! 4)} \frac1{\Delta x^{D-4}} }
\nonumber \\
& & \hspace{0cm} - \frac{\pi \cot(\frac{\pi D}{2}) \Gamma(D \!-\! 1)}{4 \Gamma(\frac{D}2)} 
\Bigl( \frac{a a' H^2}{4}\Bigr)^{\frac{D}2-2} + 
\frac{\Gamma(D \!-\! 1)}{\Gamma(\frac{D}2)} \Bigl( \frac{a a' H^2}{4}
\Bigr)^{\frac{D}2-2} \ln(a a')\Biggr\} , \label{tail} \qquad \\
\lefteqn{(a a')^{\frac{D}2 - 2} i\delta \Delta_B(x;x') = \frac{H^2}{4 \pi^{\frac{D}2}} 
\Biggl\{\frac{\Gamma(\frac{D}2)}{\Delta x^{D-4}} - \frac{\Gamma(D \!-\! 2)}{\Gamma(\frac{D}2)}
\Bigl( \frac{a a' H^2}{4}\Bigr)^{\frac{D}2-2} \Biggr\} , \label{Btail} } \\
\lefteqn{(a a')^{\frac{D}2 - 2} i\delta \Delta_C(x;x') = \frac{H^2}{4 \pi^{\frac{D}2}} 
\Biggl\{\frac{(\frac{D}2 \!-\! 3) \Gamma(\frac{D}2 \!-\! 1)}{\Delta x^{D-4}} + 
\frac{\Gamma(D \!-\! 3)}{\Gamma(\frac{D}2)} \Bigl( \frac{a a' H^2}{4}\Bigr)^{\frac{D}2-2} 
\Biggr\} . \label{Ctail} }
\end{eqnarray}
The $i\delta \Delta_I(x;x')$ determine the coincidence limits in dimensional regularization,
but only $i\delta \Delta_A(x;x')$ produces a nonzero tail term when $D=4$. The three
$i\Delta_{\Sigma I}(x;x')$ terms are each infinite series of less singular powers,
which vanish for $D=4$. They play no role in our analysis, but their expansions are given 
in Appendix A for completeness.

We can now identify the ``tail'' part of the graviton propagator,
\begin{equation}
i\Bigl[\mbox{}_{\mu\nu} \Delta^{\rm tail}_{\rho\sigma}\Bigr](x;x') \equiv
\Bigl[\mbox{}_{\mu\nu} T^A_{\rho\sigma}\Bigr]
\times i\delta \Delta_A(x;x') \; . \label{gravtail}
\end{equation}
The purpose of this paper is to check whether or not replacing the full graviton propagator 
by (\ref{gravtail}) gives those parts of the vacuum polarization which are responsible for
the secular enhancement of dynamical photons (\ref{gravphot}) and the logarithmic running
of the Coulomb potential (\ref{gravCoul}). 

\subsection{Representing Vacuum Polarization}

The one graviton loop contribution to the vacuum polarization can be expressed in terms
of expectation values of variations of the action,
\begin{eqnarray}
\lefteqn{i\Bigl[\mbox{}^{\mu} \Pi^{\nu}\Bigr](x;x') = \Biggl\langle \Omega \Biggl\vert 
\Biggl[\frac{i \delta S}{\delta A_{\mu}(x)} \Biggr]_{h A} \times \Biggl[
\frac{i \delta S}{\delta A_{\nu}(x')} \Biggr]_{h A} \Biggr\vert \Omega \Biggr\rangle }
\nonumber \\
& & \hspace{6cm} + \Biggl\langle \Omega \Biggl\vert \Biggl[
\frac{i \delta^2 S}{\delta A_{\mu}(x) \delta A_{\nu}(x')} \Biggr]_{h h} \Biggr\vert \Omega 
\Biggr\rangle \; . \label{vacpolop} \qquad
\end{eqnarray}
The subscripts $h A$ and $h h$ indicate that the operator in square brackets is to be 
expanded to that order in the weak fields $h_{\mu\nu}$ and $A_{\mu}$. Expression 
(\ref{vacpolop}) is ideal for our study because each of these two expectation values is 
{\it separately} transverse, and for {\it any} graviton field. 

The tensor structure of the de Sitter background vacuum polarization can be represented 
using two structure functions \cite{Prokopec:2002uw,Leonard:2012si,Leonard:2012ex},
\begin{equation}
i\Bigl[\mbox{}^{\mu} \Pi^{\nu}\Bigr](x;x') = \Bigl(\eta^{\mu\nu} \eta^{\rho\sigma} \!-\!
\eta^{\mu\sigma} \eta^{\nu\rho}\Bigr) \partial_{\rho} \partial'_{\sigma} F(x;x') +
\Bigl(\overline{\eta}^{\mu\nu} \overline{\eta}^{\rho\sigma} \!-\! \overline{\eta}^{\mu\sigma} 
\overline{\eta}^{\nu\rho}\Bigr) \partial_{\rho} \partial'_{\sigma} G(x;x') .
\label{vacpolform}
\end{equation}
Each of the two terms on the right hand side of (\ref{vacpolform}) is transverse so we 
can work out contributions to $F(x;x')$ and $G(x;x')$ separately, from each of the two 
expectation values in (\ref{vacpolop}), and from any part of the graviton propagator such 
as (\ref{gravtail}). Given a transverse contribution to $i[\mbox{}^{\mu} \Pi^{\nu}](x;x')$, 
the corresponding contributions to the structure functions can be inferred from selected 
components \cite{Leonard:2012ex},
\begin{eqnarray}
i\Bigl[\mbox{}^{0} \Pi^{0}\Bigr](x;x') & = & -\vec{\nabla} \!\cdot\! \vec{\nabla}' 
F(x;x') \; , \label{Fpick} \\
\eta_{\mu\nu} \!\times\! i\Bigl[\mbox{}^{\mu} \Pi^{\nu}\Bigr](x;x') & = & (D \!-\! 1)
\partial \!\cdot\! \partial' F(x;x') + (D \!-\! 2) \vec{\nabla} \!\cdot\! \vec{\nabla}' 
G(x;x') \; . \label{Gpick}
\end{eqnarray}
The same considerations imply that the two relevant counterterms (\ref{DLag}) make the 
following contributions \cite{Leonard:2013xsa},
\begin{equation}
\Delta F(x;x') = 4 \overline{C} H^2 a^{D-4} i\delta^D(x \!-\! x') \;\; , \;\;
\Delta G(x;x') = 4 \Delta C H^2 a^{D-4} i\delta^D(x \!-\! x') \; . \label{DFG}
\end{equation}

The full one loop vacuum polarization \cite{Leonard:2013xsa} contains some parts
which are de Sitter-ized versions of the flat space result \cite{Leonard:2012fs}.
However, the secular enhancement of dynamical photons (\ref{gravphot}) and the
logarithmic running of the Coulomb potential (\ref{gravCoul}) originate in the
intrisically de Sitter portions of the structure functions,
\begin{eqnarray}
F_{\rm dS}(x;x') & \!\!\! = \!\!\! & \frac{\kappa^2 H^2}{(2 \pi)^4} 
\Biggl\{ 2 \pi^2 \ln(a) i \delta^4(x \!-\! x') + \frac14 \partial^2  \Biggl[ 
\frac{\ln(\frac14 H^2 \Delta x^2)}{\Delta x^2} \Biggr] \nonumber \\
& & \hspace{5.7cm} + \partial_0^2\Biggl[ 
\frac{\ln(\frac14 H^2 \Delta x^2) \!+\! 2}{\Delta x^2} \Biggr] \Biggr\} , 
\qquad \label{FdS} \\
G_{\rm dS}(x;x') & \!\!\! = \!\!\! & \frac{\kappa^2 H^2}{(2 \pi)^4} 
\Biggl\{ -\frac83 \pi^2 \ln(a) i \delta^4(x \!-\! x') - \frac13 \partial^2 
\Biggl[ \frac{\ln(\frac14 H^2 \Delta x^2)}{\Delta x^2} \Biggr] \Biggr\} .
\label{GdS}
\end{eqnarray}
The enhancement of dynamical photons actually derives entirely from just the 
$\ln(a)$ part of $F_{dS}(x;x')$ \cite{Wang:2014tza}. In contrast, all terms 
on the first lines of (\ref{FdS}-\ref{GdS}) contribute to the logarithmic 
running of the Coulomb potential \cite{Glavan:2013jca}. The terms on the 
second line of expression (\ref{FdS}) do not contribute to either the 
enhancement of photons or the running of the Coulomb potential.

\section{Vacuum Polarization from the Tail}

\begin{figure}[ht]
\center
\includegraphics[]{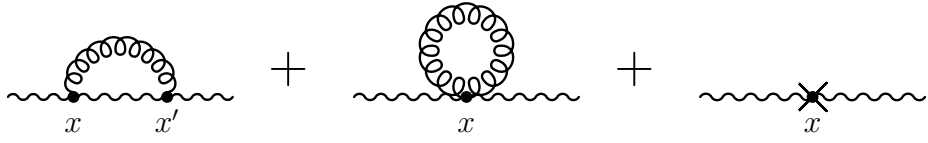}
\caption{\label{fig:photon} Feynman diagrams relevant to the
one loop vacuum polarization from gravitons. Wavy lines are photons,
curly lines are gravitons and the cross represents counterterms.}
\label{vacpolgraphs}
\end{figure}

This section presents the key computation of the tail contribution to the
two structure functions of the vacuum polarization. Because each of the 
terms in the operator expression (\ref{vacpolop}) is separately transverse,
as is the contribution from the counter-action, we derive separate results 
for each of the three diagrams in Figure~\ref{vacpolgraphs}. Because the 
counterterms contribute at leading logarithm order it is necessary to 
retain dimensional regularization until the end. (The same thing was found
in deriving the tail contribution to the fermion wave function 
\cite{Miao:2008sp}.) However, extensive simplifications result from 
anticipating terms which must vanish in the renormalized, unregulated 
limit. We begin with the simple 4-point contribution, then proceed to the
more complicated contribution from two 3-point vertices, and finally add
the appropriate counterterms.

\subsection{The 4-point contribution}

The primitive 4-point contribution is the middle diagram of 
Fig.~\ref{vacpolgraphs} and has the operator representation,
\begin{equation}
i\Bigl[\mbox{}^{\mu} \Pi_{\rm 4pt}^{\nu}\Bigr](x;x') = \partial_{\rho} 
\partial'_{\sigma} \Bigl\langle \Omega \Bigl\vert a^{D-4} 
\sqrt{-\widetilde{g}} \, \Bigl( \widetilde{g}^{\mu\sigma} \widetilde{g}^{\nu\rho} 
\!-\! \widetilde{g}^{\mu\nu} \widetilde{g}^{\rho\sigma} \Bigr) 
i \delta^D(x \!-\! x') \Bigr\vert \Omega \Bigr\rangle_{hh} \; . \label{4ptop}
\end{equation}
This expression is exact. Because the tail contribution comes from the purely
spatial components of the graviton field we can use relation (\ref{Fpick}) to 
write a simple relation for the tail part of the structure function $F(x;x')$,
\begin{equation}
-\vec{\nabla} \!\cdot\! \vec{\nabla}' F_{\rm 4t}(x;x') = \partial_i 
\partial'_j \Bigl\langle \Omega \Bigl\vert a^{D-4} \sqrt{-\widetilde{g}} \, 
\widetilde{g}^{ij} i \delta^D(x \!-\! x') \Bigr\vert \Omega \Bigr\rangle_{\rm tail} 
\; . \qquad \label{F4top1} 
\end{equation}
Isotropy implies,
\begin{eqnarray}
F_{\rm 4t}(x;x') & = & -\frac1{D\!-\!1} \Bigl\langle \Omega \Bigl\vert a^{D-4} 
\sqrt{-\widetilde{g}} \, \widetilde{g}^{kk} i \delta^D(x \!-\! x') \Bigr\vert 
\Omega \Bigr\rangle_{\rm tail} \; , \qquad \label{F4top2} \\ 
& = & \frac14 D (D \!-\! 5) \kappa^2 a^{D-4} i\delta \Delta_A(x;x) 
i \delta^D(x \!-\! x') \; . \qquad \label{F4t}
\end{eqnarray}
Expression (\ref{F4t}) agrees with the result (66) reported in 
\cite{Leonard:2013xsa}.

Relation (\ref{Gpick}) determines the structure function $G(x;x')$,
\begin{eqnarray}
\lefteqn{ (D\!-\!1) \partial \!\cdot\! \partial' F_{\rm 4t} \!+\! 
(D \!-\! 2) \vec{\nabla} \!\cdot\! \vec{\nabla}' G_{\rm 4t} = \partial_0 
\partial'_0 \Bigl\langle \Omega \Bigl\vert a^{D-4} \sqrt{-\widetilde{g}} \, 
\widetilde{g}^{kk} i \delta^D(x \!-\! x') \Bigr\vert \Omega 
\Bigr\rangle_{\rm tail} } \nonumber \\
& & \hspace{1.3cm} + \partial_i \partial'_j \Bigl\langle \Omega \Bigl\vert 
a^{D-4} \sqrt{-\widetilde{g}} \, \Bigl(\widetilde{g}^{ik} \widetilde{g}^{jk} \!+\!
\widetilde{g}^{ij} (1 \!-\! \widetilde{g}^{kk}) \Bigr) i \delta^D(x \!-\! x') 
\Bigr\vert \Omega \Bigr\rangle_{\rm tail} \; . \qquad \label{G4t1}
\end{eqnarray}
Using relation (\ref{F4top2}) and exploiting isotropy implies,
\begin{eqnarray}
G_{\rm 4t} & \!\!\!=\!\!\! & \Bigl\langle \Omega \Bigl\vert \frac{a^{D-4} 
\sqrt{-\widetilde{g}}}{(D\!-\!1) (D\!-\!2)} \Bigl(\widetilde{g}^{k\ell} 
\widetilde{g}^{k\ell} \!+\! \widetilde{g}^{kk} [(D \!-\! 2) \!-\! 
\widetilde{g}^{\ell\ell}] \Bigr) i \delta^D(x \!-\! x') \Bigr\vert \Omega 
\Bigr\rangle_{tail} \label{G4top} \; , \qquad \\
& \!\!\!=\!\!\! & -\Bigl[D - \Bigl(\frac{D \!-\! 1}{D \!-\! 3}\Bigr)\Bigr] 
\kappa^2 a^{D-4} i\delta \Delta_A(x;x) i \delta^D(x \!-\! x') \; . 
\label{G4t} \qquad
\end{eqnarray}
Expression (\ref{G4t}) agrees with the result (67) reported in 
\cite{Leonard:2013xsa}. 

\subsection{The 3-point contribution}

The primitive 3-point contribution is the left hand diagram of 
Fig.~\ref{vacpolgraphs}. From the first term of the operator expression 
(\ref{vacpolop}) we can infer a simpler operator expression for it,
\begin{eqnarray}
\lefteqn{i\Bigl[\mbox{}^{\mu} \Pi_{\rm 3pt}^{\nu}\Bigr](x;x') = -\partial_{\rho} 
\partial'_{\sigma} \Biggl\{ \Bigl\langle \Omega \Bigl\vert
\Bigl[ \sqrt{-\widetilde{g}} \, \widetilde{g}^{\rho [\alpha} 
\widetilde{g}^{\beta ] \mu} \Bigr]_{h(x)} \!\times\! \Bigl[ \sqrt{-\widetilde{g}} 
\, \widetilde{g}^{\sigma [\gamma} \widetilde{g}^{\delta ] \nu} \Bigr]_{h(x')}
\Bigr\vert \Omega \Bigr\rangle } \nonumber \\ 
& & \hspace{6.5cm} \times 4 (a a')^{D-4} \partial_{\alpha} \partial'_{\gamma}
i\Bigl[\mbox{}_{\beta} \Delta_{\delta}\Bigr](x;x') \Biggr\} \; ,
\label{3top1} \qquad 
\end{eqnarray}
where square bracketed indices are anti-symmetrized. If we specialize to just
the tail contribution then the expectation value on the first line of (\ref{3top1})
goes like $1/\Delta x^{D-4}$. Hence the entire curly-bracketed term is at most
logarithmically divergent, and that only when both of the derivatives on the
second line of (\ref{3top1}) act on the most singular part of the photon
propagator (\ref{photprop}). Because the less singular parts vanish for $D = 4$
we can make the simplification,
\begin{equation}
4 ( a a')^{D-4} \partial_{\alpha} \partial'_{\gamma} i\Bigl[\mbox{}_{\beta}
\Delta_{\delta}\Bigr](x;x') \longrightarrow 4 (a a')^{\frac{D}2 - 2} 
\eta_{\beta\delta} \partial_{\alpha} \partial'_{\gamma} i\Delta(x;x') \; .
\label{photsimp}
\end{equation}

Substituting (\ref{photsimp}) in (\ref{3top1}), and exploiting relation 
(\ref{Fpick}), gives an operator expression for the tail contribution to the 
$F(x;x')$ structure function,
\begin{eqnarray}
\lefteqn{-\vec{\nabla} \!\cdot\! \vec{\nabla}' F_{3t}(x;x') = -\partial_i \partial'_j 
\Biggl\{ \Bigl\langle \Omega \Bigl\vert \Bigl[ \sqrt{-\widetilde{g}} \, 
\widetilde{g}^{ik} \Bigr]_{h(x)} \!\times\! \Bigl[ \sqrt{-\widetilde{g}} 
\, \widetilde{g}^{i\ell} \Bigr]_{h(x')} \Bigr\vert \Omega \Bigr\rangle_{\rm tail} } 
\nonumber \\ 
& & \hspace{5.2cm} \times (a a')^{\frac{D}2-2} \Bigl[ \delta_{k\ell} \partial_0 
\partial'_0 \!-\! \partial_k \partial'_{\ell}\Bigr] i\Delta(x;x') \Biggr\} \; , 
\label{F3top1} \qquad \\
& & \hspace{0cm} = -\kappa^2 \partial_i \partial'_j \Biggl\{ \Bigl\langle \Omega 
\Bigl\vert \frac14 h^2 \delta^{ik} \delta^{j\ell} \!-\! \frac12 h^{ik} h \delta^{j\ell}
\!-\! \frac12 h \delta^{ik} h^{j\ell} \!+\! h^{ik} h^{j\ell} \Bigr\vert \Omega 
\Bigr\rangle_{\rm tail} \nonumber \\
& & \hspace{5.2cm} \times (a a')^{\frac{D}2-2} \Bigl[ \delta_{k\ell} \partial_0 
\partial'_0 \!-\! \partial_k \partial'_{\ell}\Bigr] i\Delta(x;x') \Biggr\} \; . 
\label{F3top2}
\end{eqnarray}
Substituting the tail part of the propagator (\ref{gravtail}) and performing 
the simple contractions implies,
\begin{equation}
F_{3t}(x;x') = \kappa^2 i\delta \Delta_A(x;x') (a a')^{\frac{D}2-2} 
\Bigl[ (D \!-\! 1) \partial_0 \partial'_0 \!-\! \vec{\nabla} \!\cdot\! \vec{\nabla}'
\Bigr] i\Delta(x;x') \; . \label{F3prop}
\end{equation}
The final step is extracting the derivatives from inside the square brackets of
(\ref{F3prop}), which is done generically in Appendix B. From relation (\ref{keyID})
we infer,
\begin{eqnarray}
\lefteqn{F_{3t}(x;x') = -\frac{\kappa^2 H^2 \partial \!\cdot\! \partial'}{64 \pi^4} 
\Biggl[\frac{\ln(\frac14 H^2 \Delta x^2) \!-\! 4}{\Delta x^2}\Biggr]  
-\frac{\kappa^2 H^2 \partial_0 \partial'_0}{16 \pi^4} \Biggl[
\frac{\ln(\frac14 H^2 \Delta x^2) \!+\! 2}{\Delta x^2}\Biggr] } \nonumber \\
& & \hspace{5cm} -\frac{\kappa^2 H^{D-2} (D \!-\! 1) \Gamma(\frac{D}2 \!+\! 1) \, 
i\delta^D(x \!-\! x')}{ (4 \pi)^{\frac{D}2} (D \!-\!3) (D \!-\! 4)} \; . \qquad
\label{F3final}
\end{eqnarray}
Both the divergence and the $\ln(\frac14 H^2 \Delta x^2)$ terms agree with the
results reported in equations (129) and (130) of \cite{Leonard:2013xsa}.

Relations (\ref{Fpick}-\ref{Gpick}) provide an operator expression for the
$G(x;x')$ structure function,
\begin{equation}
(D \!-\! 1) \partial \!\cdot\! \partial' F(x;x') + (D \!-\! 2)
\vec{\nabla} \!\cdot\! \vec{\nabla}' G(x;x') = \vec{\nabla} \!\cdot\! 
\vec{\nabla}' F(x;x') + i \Bigl[\mbox{}^{k} \Pi^{k}\Bigr](x;x') \; .
\label{Gop}
\end{equation}
Specializing (\ref{Gop}) to the 3-point tail contribution gives,
\begin{eqnarray}
\lefteqn{ (D \!-\! 2) \vec{\nabla} \!\cdot\! \vec{\nabla}' G_{\rm 3t}(x;x') =
-(D \!-\! 2) \vec{\nabla} \!\cdot\! \vec{\nabla}' F_{\rm 3t}(x;x') + (D \!-\!1)
\partial_0 \partial'_0 F_{\rm 3t}(x;x') } \nonumber \\
& & \hspace{-.5cm} -\partial_{\rho} \partial'_{\sigma} \Biggl\{ \Bigl\langle 
\Omega \Bigl\vert \Bigl[ \sqrt{-\widetilde{g}} \Bigl( \widetilde{g}^{\rho \alpha} 
\widetilde{g}^{\beta\mu} \!-\! \widetilde{g}^{\rho\beta} \widetilde{g}^{\alpha\mu}
\Bigr)\Bigr]_{h(x)} \!\times\! \Bigl[ \sqrt{-\widetilde{g}} \Bigl( 
\widetilde{g}^{\sigma\gamma} \widetilde{g}^{\delta\nu} \!-\! 
\widetilde{g}^{\sigma\delta} \widetilde{g}^{\gamma\nu}\Bigr) \Bigr]_{h(x')}
\Bigr\vert \Omega \Bigr\rangle_{\rm tail} \nonumber \\
& & \hspace{6.5cm} \times (a a')^{D-4} \eta_{\beta\delta} \partial_{\alpha} 
\partial'_{\gamma} i\Delta(x;x') \Biggr\} \; . \label{3tGop1} \qquad 
\end{eqnarray}
The $\rho = \sigma = 0$ component of the contraction in (\ref{3tGop1}) cancels 
the factor of $(D-1) \partial_0 \partial'_0 F_{3t}(x;x')$. Expanding out the
remaining terms gives,
\begin{eqnarray}
\lefteqn{ (D \!-\! 2) \vec{\nabla} \!\cdot\! \vec{\nabla}' G_{\rm 3t}(x;x') =
-(D \!-\! 2) \vec{\nabla} \!\cdot\! \vec{\nabla}' F_{\rm 3t}(x;x') } \nonumber \\
& & \hspace{-.5cm} + \partial_{0} \partial'_{i} \Biggl\{ \Bigl\langle \Omega \Bigl\vert 
\sqrt{-\widetilde{g}} \, \widetilde{g}^{k\ell} \!\times\! \sqrt{-\widetilde{g}} \Bigl( 
\widetilde{g}^{ij} \widetilde{g}^{k\ell} \!-\! \widetilde{g}^{i\ell} \widetilde{g}^{jk}
\Bigr) \Bigr\vert \Omega \Bigr\rangle_{\rm tail} (a a')^{\frac{D}2-2} \partial_0 
\partial'_j i\Delta(x;x') \Biggr\} \nonumber \\
& & \hspace{-.5cm} + \partial_{i} \partial'_{0} \Biggl\{ \Bigl\langle \Omega \Bigl\vert 
\sqrt{-\widetilde{g}} \Bigl( \widetilde{g}^{ij} \widetilde{g}^{k\ell} \!-\! 
\widetilde{g}^{i\ell} \widetilde{g}^{jk} \Bigr) \!\times\! \sqrt{-\widetilde{g}} \,
\widetilde{g}^{k\ell} \Bigr\vert \Omega \Bigr\rangle_{\rm tail} (a a')^{\frac{D}2-2} 
\partial_j \partial'_0 i\Delta(x;x') \Biggr\} \nonumber \\
& & \hspace{-.5cm} - \partial_{i} \partial'_{j} \Biggl\{ \Bigl\langle \Omega \Bigl\vert 
\sqrt{-\widetilde{g}} \Bigl( \widetilde{g}^{im} \widetilde{g}^{k\ell} \!-\! 
\widetilde{g}^{i\ell} \widetilde{g}^{mk} \Bigr) \!\times\! \sqrt{-\widetilde{g}}
\Bigl(\widetilde{g}^{jn} \widetilde{g}^{k\ell} \!-\! \widetilde{g}^{j\ell}
\widetilde{g}^{kn}\Bigr) \Bigr\vert \Omega \Bigr\rangle_{\rm tail} \nonumber \\
& & \hspace{7cm} \times (a a')^{\frac{D}2-2} \partial_m \partial'_n i\Delta(x;x') 
\Biggr\} , \label{3tGop2} \qquad \\
& & \hspace{-.7cm} = -\kappa^2 \vec{\nabla} \!\cdot\! \vec{\nabla}' \Biggl\{ \!
(a a')^{\frac{D}2-2} i\delta \Delta_A(x;x') \! \Biggl[ 2 \Bigl(
\frac{D^2 \!-\! 5D \!+\! 5}{D \!-\! 3}\Bigr) \vec{\nabla} \!\cdot\! \vec{\nabla}'
\!+\! (D \!-\! 2) (D \!-\! 1) \partial_0 \partial'_0 \Biggr] \nonumber \\
& & \hspace{0cm} \times i\Delta(x;x') \Biggr\} + (D\!-\!2)^2 \kappa^2 \partial_0 
\partial'_i \Biggl\{ (a a')^{\frac{D}2-2} i\delta \Delta_A(x;x') \partial_0 
\partial'_i i\Delta(x;x') \Biggr\} \nonumber \\
& & \hspace{1cm} + (D\!-\!2)^2 \kappa^2 \partial_i \partial'_0 \Biggl\{
(a a')^{\frac{D}2-2} i\delta \Delta_A(x;x') \partial_i \partial'_0 i\Delta(x;x') 
\Biggr\} \nonumber \\
& & \hspace{1.5cm} - (D\!-\!4) (D \!-\! 1) \kappa^2 \partial_i \partial'_j \Biggl\{
(a a')^{\frac{D}2-2} i\delta \Delta_A(x;x') \partial_i \partial'_j i\Delta(x;x') 
\Biggr\} , \label{3tG1} \qquad
\end{eqnarray}
where some of the terms from the first line of (\ref{3tG1}) derive from the operator
expressions on the last line of (\ref{3tGop2}) and spatial translation invariance has
been exploited. It remains to extract the inner derivatives using relation 
(\ref{keyID}) and solve for $G_{\rm 3t}(x;x')$,
\begin{equation}
G_{3t}(x;x') = \frac{\kappa^2 H^2 \partial \!\cdot\! \partial'}{32 \pi^4} \Biggl[
\frac{\ln(\frac14 H^2 \Delta x^2) \!+\! 2}{\Delta x^2} \Biggr] \; . \label{G3final}
\end{equation}
This result agrees with the $\ln(\frac14 H^2 \Delta x^2)$ term reported in 
equation (132) of \cite{Leonard:2013xsa}. However, it has neither the ultraviolet
divergence reported in equation (131) of that paper, nor the associated factor of 
$\ln(\mu^2 \Delta x^2)$ reported in equation (132). These terms come from the 
non-tail part of the graviton propagator. 

\subsection{Tail Renormalization}

The right hand diagram of Fig.~\ref{vacpolgraphs} stands for renormalization
counterterms. Their contributions to the two structure functions was given in
equation (\ref{DFG}). We must bear in mind the fact that the coefficients 
$\overline{C}$ and $\Delta C$ are not those appropriate to the full vacuum 
polarization \cite{Leonard:2013xsa} but rather just the parts needed to cancel
the divergences in our tail results (\ref{F4t}) and (\ref{F3final}) for $F(x;x')$
and (\ref{G4t}) and (\ref{G3final}) for $G(x;x)$. 

Based on expressions (\ref{F4t}) and (\ref{F3final}) the best choice for the
$\overline{C}$ counterterm is,
\begin{equation}
\overline{C} = \frac{\kappa^2 H^{D-4}}{(4\pi)^{\frac{D}2}} \Biggl\{ 
\frac{D (D \!-\! 5) \Gamma(D \!-\! 1) \pi \cot(\frac{\pi D}{2})}{16 
\Gamma(\frac{D}2)} + \frac{(D \!-\! 1) \Gamma(\frac{D}2 \!+\! 1)}{4 (D \!-\!3)
(D \!-\! 4)} + 1 \Biggr\} \; . \label{Cbar}
\end{equation}
After combining with the primitive results (\ref{F4t}) and (\ref{F3final}) and
taking the unregulated limit we obtain,
\begin{eqnarray}
\lefteqn{F_{\rm tail}(x;x') = \frac{\kappa^2 H^2}{(2 \pi)^4} \Biggl\{2 \pi^2
\ln(a) i\delta^4(x \!-\! x') + \frac14 \partial^2 \Biggl[ 
\frac{\ln(\frac14 H^2 \Delta x^2)}{\Delta x^2} \Biggr] } \nonumber \\
& & \hspace{7.5cm} + \partial_0^2 \Biggl[ \frac{\ln(\frac14 H^2 \Delta x^2) 
\!+\! 2}{\Delta x^2} \Biggr] \Biggr\} . \label{Ftail} \qquad
\end{eqnarray}
Expression (\ref{Ftail}) agrees exactly with the intrinsically de Sitter part
of the full $F(x;x')$ structure function (\ref{FdS}), including even the parts
on the second line which play no role in either the secular enhancement of
dynamical photons \cite{Wang:2014tza} or the logarithmic running of the Coulomb 
potential \cite{Glavan:2013jca}.

Based on expressions (\ref{G4t}) and (\ref{G3final}) the best choice for the
nocovariant $\Delta C$ counterterm is,
\begin{equation}
\Delta C = \frac{\kappa^2 H^{D-4}}{(4\pi)^{\frac{D}2}} \Biggl\{-  
\frac{(D^2 \!-\! 4D \!+\! 1) \Gamma(D \!-\! 1) \pi \cot(\frac{\pi D}{2})}{4  
(D \!-\! 3) \Gamma(\frac{D}2)} + 1 \Biggr\} \; . \label{DeltaC}
\end{equation}
The unregulated limit of the renormalized tail contribution to $G(x;x')$ is,
\begin{equation}
G_{\rm tail}(x;x') = \frac{\kappa^2 H^2}{(2 \pi)^4} \Biggl\{-4 \pi^2
\ln(a) i\delta^4(x \!-\! x') - \frac12 \partial^2 \Biggl[ 
\frac{\ln(\frac14 H^2 \Delta x^2)}{\Delta x^2} \Biggr] \Biggr\} . 
\label{Gtail}
\end{equation}
Expression (\ref{Gtail}) does not agree with (\ref{GdS}) because the primitive 
3-point tail contribution (\ref{G3final}) lacks both the divergence and the
associated $\mu$-dependent logarithm of the full 3-point result 
\cite{Leonard:2013xsa}.

\section{Discussion}

Our aim has been to see how much of the intrinsically de Sitter part 
(\ref{FdS}-\ref{GdS}) of the vacuum polarization arises from replacing
the full graviton propagator (\ref{gravprop}) with just its tail part 
(\ref{gravtail}). Our result is that the tail reproduces all of (\ref{FdS})
but not all of (\ref{GdS}). This means that the graviton tail is responsible 
for the the secular enhancement of dynamical photons (\ref{gravphot}), but 
not for all of the logarithmic running of the Coulomb potential (\ref{gravCoul}).
The remaining parts of (\ref{GdS}) come from using the most singular part 
of the graviton propagator in the 3-point contribution. Although these terms 
have no factor of $H^2$, they do contain $\frac1{a a'} = H^2 \eta \eta'$. 
When the inner derivatives are passed through this term they can act on the 
$\eta \eta'$ and leave the required factor of $H^2$.

Our result means that the tail term is {\it not} responsible for all the 
interesting secular effects mediated by the one loop vacuum polarization.
This may not be the setback it would seem for the crucial task of extending
Satrobinsky's stochastic technique \cite{Starobinsky:1986fx,Starobinsky:1994bd}
to quantum gravity. The large logarithms of interest derive from three sources:
\begin{enumerate}
\item{Explicit factors of $\ln(a a')$ and $\ln(H^2 \Delta x^2)$ in the tail 
part of the graviton propagator (\ref{gravtail});}
\item{Factors of $(a a')^{\frac{D}2-2}/(D-4)$ and $(\Delta x)^{D-4}/(D-4)$
which arise either in primitive ultraviolet divergences or in the counterterms
which remove them; and}
\item{The integration of interaction vertices which one must do in higher 
loop diagrams.}
\end{enumerate}
The one loop tail contributions (\ref{Ftail}) and (\ref{Gtail}) that we have
computed come from the first two sources. The reason (\ref{Gtail}) does not
give all the interesting parts (\ref{GdS}) of the $G(x;x')$ structure function
is that we have missed some ultraviolet divergences from the most singular 
part of the propagator. {\it These sorts of terms are easy to recover using
renormalization group techniques.} The ``hard'' contributions --- the ones for
which one loop divergences do not predict higher loop results --- are those
from the other two sources. So perhaps the key to dealing with the large
logarithms is to combine Starobinsky's technique with the renormalization
group.

\vskip 1cm

\centerline{\bf Acknowledgements}

We are grateful for conversations and correspondence with Y. Z Chu, N. C.
Tsamis and C. L. Wang. This work was partially supported by Taiwan MOST 
grants 103-2112-M-006-001-MY3 and 106-2112-M-006-008-; by the D-ITP 
consortium, a program of the Netherlands Organization for Scientific 
Research (NWO) that is funded by the Dutch Ministry of Education, 
Culture and Science (OCW); by NSF grant PHY-1506513; and by the 
Institute for Fundamental Theory at the University of Florida.

\section{Appendix A: $i\Delta_{\Sigma I}(x;x')$ Expansions}

The infinite series expansions for the scalar propagators (\ref{scalarprops}) are:
\begin{eqnarray}
\lefteqn{ i\Delta_{\Sigma A}(x;x') = \frac{H^{D-2}}{(4\pi)^{\frac{D}2}} 
\sum_{n=1}^{\infty} \Bigl( \frac{ a a' H^2 \Delta x^2}{4}\Bigr)^n } \nonumber \\
& & \hspace{2cm} \times \Biggl\{ 
\frac{\Gamma(n \!+\! D \!-\! 1)}{n \, \Gamma(n \!+\! \frac{D}2)} - 
\frac{ \Gamma(n \!+\! \frac{D}2 \!+\! 1)}{(n \!-\! \frac{D}2 \!+\! 2) 
(n \!+\! 1)!} \Bigl( \frac{4}{a a' H^2 \Delta x^2} \Bigr)^{\frac{D}2-2} 
\Biggr\} , \qquad \label{DSigmaA} \\
\lefteqn{ i\Delta_{\Sigma B}(x;x') = \frac{H^{D-2}}{(4\pi)^{\frac{D}2}} 
\sum_{n=1}^{\infty} \Bigl( \frac{ a a' H^2 \Delta x^2}{4}\Bigr)^n } \nonumber \\
& & \hspace{3.5cm} \times \Biggl\{ 
\frac{\Gamma(n \!+\! D \!-\! 2)}{\Gamma(n \!+\! \frac{D}2)} - 
\frac{ \Gamma(n \!+\! \frac{D}2)}{ (n \!+\! 1)!} 
\Bigl( \frac{4}{a a' H^2 \Delta x^2} \Bigr)^{\frac{D}2-2} 
\Biggr\} , \qquad \label{DSigmaB} \\
\lefteqn{ i\Delta_{\Sigma C}(x;x') = \frac{H^{D-2}}{(4\pi)^{\frac{D}2}} 
\sum_{n=1}^{\infty} \Bigl( \frac{ a a' H^2 \Delta x^2}{4}\Bigr)^n } \nonumber \\
& & \hspace{0cm} \times \Biggl\{ 
\frac{(n \!+\! 1) \Gamma(n \!+\! D \!-\! 3)}{\Gamma(n \!+\! \frac{D}2)} - 
\frac{ (n \!-\! \frac{D}2 \!+\! 3) \Gamma(n \!+\! \frac{D}2 \!-\! 1)}{ (n \!+\! 1)!} 
\Bigl( \frac{4}{a a' H^2 \Delta x^2} \Bigr)^{\frac{D}2-2} 
\Biggr\} . \qquad \label{DSigmaC}
\end{eqnarray}

\section{Appendix B: Extracting Derivatives}

Evaluating the 3-point contributions requires that we wish pass derivatives of the 
photon propagator to the left of $(a a')^{\frac{D}2-2} i\delta \Delta_A(x;x')$ in 
expressions of the form,
\begin{equation}
(a a')^{\frac{D}2 -2} i\delta \Delta_A(x;x') \partial_{\mu} \partial'_{\nu} 
i\Delta(x;x') \; .
\end{equation}
The propagator $i\Delta(x;x')$ goes like $1/\Delta x^{D-2}$. From equation (\ref{tail}) 
we see that $(a a')^{\frac{D}2-2} i\delta \Delta_A(x;x')$ contains three distinct sorts 
of coordinate dependence. The result passing derivatives through each of these terms
is,
\begin{eqnarray}
\frac1{\Delta x^{D-4}} \, \partial_{\mu} \partial'_{\nu} \frac1{\Delta x^{D-2}} & = &
\frac{[D \partial_{\mu} \partial'_{\nu} \!-\! \eta_{\mu\nu} \partial \!\cdot\! \partial'
]}{4 (D \!-\! 3)} \, \frac1{\Delta x^{2D-6}} \; , \label{relation1} \\
(a a')^{\frac{D}2-2} \, \partial_{\mu} \partial'_{\nu} \frac1{\Delta x^{D-2}} & = &
\Bigl[ \partial_{\mu} \!-\! \Bigl(\frac{D}2 \!-\! 2\Bigr) H a \delta^0_{~\mu}\Bigr]
\nonumber \\
& & \hspace{1cm} \times \Bigl[ \partial'_{\nu} \!-\! \Bigl(\frac{D}2 \!-\! 2\Bigr) H a' 
\delta^0_{~\nu}\Bigr] \Bigl[ \frac{(a a')^{\frac{D}2-2}}{\Delta x^{D-2}} \Bigr] \; , 
\label{relation2} \qquad \\
\ln(a a') \, \partial_{\mu} \partial'_{\nu} \frac1{\Delta x^{D-2}} & = &
\partial_{\mu} \partial'_{\nu} \Bigl[ \frac{\ln(a a')}{\Delta x^{D-2}} \Bigr] 
- \partial_{\mu} \Bigl[ \frac{H a' \delta^0_{~\nu}}{\Delta x^{D-2}} 
\Bigr] - \partial'_{\nu} \Bigl[ \frac{H a \delta^0_{~\mu}}{\Delta x^{D-2}} \Bigr] \; . 
\label{relation3} \qquad 
\end{eqnarray}
The first terms on the right hand side of relations (\ref{relation1}-\ref{relation3})
give the derivatives acting on the product $(a a')^{\frac{D}2-2} i\delta \Delta_A(x;x')
i\Delta(x;x')$. That product is integrable for $D=4$ so we can take its unregulated 
limit. The secondary terms of relations (\ref{relation2}-\ref{relation3}) cancel in $D=4$
dimensions, so it only remains to consider the second term on the right of relation
(\ref{relation1}),
\begin{eqnarray}
\lefteqn{\partial \!\cdot\! \partial' \frac1{\Delta x^{2D-6}} = \partial \!\cdot\! 
\partial' \Biggl[ \frac1{\Delta x^{2D-6}} \!-\! \frac{\mu^{D-4}}{\Delta x^{D-2}}\Biggr]
- \frac{4 \pi^{\frac{D}2} \mu^{D-4} i\delta^D(x \!-\! x')}{\Gamma(\frac{D}2 \!-\! 1)}
\; , } \\
& & \hspace{-.5cm} = -\Bigl( \frac{D \!-\! 4}{2}\Bigr) \partial \!\cdot\! \partial' \Biggl[ 
\frac{\ln(\mu^2 \Delta x^2)}{\Delta x^{2}} \Biggr] + O\Bigl( (D \!-\! 4)^2\Bigr) - 
\frac{4 \pi^{\frac{D}2} \mu^{D-4} i\delta^D(x \!-\! x')}{\Gamma(\frac{D}2 \!-\! 1)} 
\; . \qquad
\end{eqnarray}
Setting $\mu = \frac12 H$ and putting everything together gives,
\begin{eqnarray}
\lefteqn{ (a a')^{\frac{D}2 -2} i\delta \Delta_A(x;x') \partial_{\mu} \partial'_{\nu} 
i\Delta(x;x') = -\frac{H^2 \partial_{\mu} \partial'_{\nu}}{32 \pi^4} \Biggl[
\frac{ \ln(\frac14 H^2 \Delta x^2) \!+\! 2}{\Delta x^2}\Biggr] } \nonumber \\
& & \hspace{-.5cm} + \frac{H^2 \eta_{\mu\nu} \partial \!\cdot\! \partial'}{128 \pi^4}
\Biggl[ \frac{\ln(\frac14 H^2 \Delta x^2)}{\Delta x^2} \Biggr] \!+\! 
\frac{H^{D-2} \eta_{\mu\nu}}{(4\pi)^{\frac{D}2}} \frac{ \Gamma(\frac{D}2 \!+\! 1) \, 
i\delta^D(x \!-\! x')}{2 (D \!-\! 3) (D \!-\! 4)} \!+\! O(D \!-\! 4) . 
\label{keyID} \qquad
\end{eqnarray}


\begin{thebibliography}{99}

%\cite{Lifshitz:1945du}
\bibitem{Lifshitz:1945du} 
  E.~Lifshitz,
  %``Republication of: On the gravitational stability of the expanding universe,''
  J.\ Phys.\ (USSR) {\bf 10}, 116 (1946)
  [Gen.\ Rel.\ Grav.\  {\bf 49}, no. 2, 18 (2017)].
  doi:10.1007/s10714-016-2165-8
  %%CITATION = doi:10.1007/s10714-016-2165-8;%%
  %280 citations counted in INSPIRE as of 08 Apr 2018
  
%\cite{Grishchuk:1974ny}
\bibitem{Grishchuk:1974ny} 
  L.~P.~Grishchuk,
  %``Amplification of gravitational waves in an istropic universe,''
  Sov.\ Phys.\ JETP {\bf 40}, 409 (1975)
  [Zh.\ Eksp.\ Teor.\ Fiz.\  {\bf 67}, 825 (1974)].
  %%CITATION = SPHJA,40,409;%%
  %647 citations counted in INSPIRE as of 08 Apr 2018
  
%\cite{Ford:1977dj}
\bibitem{Ford:1977dj} 
  L.~H.~Ford and L.~Parker,
  %``Quantized Gravitational Wave Perturbations in Robertson-Walker Universes,''
  Phys.\ Rev.\ D {\bf 16}, 1601 (1977).
  doi:10.1103/PhysRevD.16.1601
  %%CITATION = doi:10.1103/PhysRevD.16.1601;%%
  %166 citations counted in INSPIRE as of 08 Apr 2018  
  
%\cite{Starobinsky:1979ty}
\bibitem{Starobinsky:1979ty} 
  A.~A.~Starobinsky,
  %``Spectrum of relict gravitational radiation and the early state of the universe,''
  JETP Lett.\  {\bf 30}, 682 (1979)
  [Pisma Zh.\ Eksp.\ Teor.\ Fiz.\  {\bf 30}, 719 (1979)].
  %%CITATION = JTPLA,30,682;%%
  %1275 citations counted in INSPIRE as of 08 Apr 2018    

%\cite{DeWitt:1960fc}
\bibitem{DeWitt:1960fc} 
  B.~S.~DeWitt and R.~W.~Brehme,
  %``Radiation damping in a gravitational field,''
  Annals Phys.\  {\bf 9}, 220 (1960).
  doi:10.1016/0003-4916(60)90030-0
  %%CITATION = doi:10.1016/0003-4916(60)90030-0;%%
  %512 citations counted in INSPIRE as of 08 Apr 2018
  
%\cite{Chu:2011ip}
\bibitem{Chu:2011ip} 
  Y.~Z.~Chu and G.~D.~Starkman,
  %``Retarded Green's Functions In Perturbed Spacetimes For Cosmology and Gravitational Physics,''
  Phys.\ Rev.\ D {\bf 84}, 124020 (2011)
  doi:10.1103/PhysRevD.84.124020
  [arXiv:1108.1825 [astro-ph.CO]].
  %%CITATION = doi:10.1103/PhysRevD.84.124020;%%
  %13 citations counted in INSPIRE as of 08 Apr 2018
  
%\cite{Tanaka:1996ht}
\bibitem{Tanaka:1996ht} 
  T.~Tanaka, Y.~Mino, M.~Sasaki and M.~Shibata,
  %``Gravitational waves from a spinning particle in circular orbits around a rotating black hole,''
  Phys.\ Rev.\ D {\bf 54}, 3762 (1996)
  doi:10.1103/PhysRevD.54.3762
  [gr-qc/9602038].
  %%CITATION = doi:10.1103/PhysRevD.54.3762;%%
  %63 citations counted in INSPIRE as of 08 Apr 2018 
  
%\cite{Mino:1996nk}
\bibitem{Mino:1996nk} 
  Y.~Mino, M.~Sasaki and T.~Tanaka,
  %``Gravitational radiation reaction to a particle motion,''
  Phys.\ Rev.\ D {\bf 55}, 3457 (1997)
  doi:10.1103/PhysRevD.55.3457
  [gr-qc/9606018].
  %%CITATION = doi:10.1103/PhysRevD.55.3457;%%
  %330 citations counted in INSPIRE as of 08 Apr 2018

%\cite{Quinn:1996am}
\bibitem{Quinn:1996am} 
  T.~C.~Quinn and R.~M.~Wald,
  %``An Axiomatic approach to electromagnetic and gravitational radiation reaction of particles in curved space-time,''
  Phys.\ Rev.\ D {\bf 56}, 3381 (1997)
  doi:10.1103/PhysRevD.56.3381
  [gr-qc/9610053].
  %%CITATION = doi:10.1103/PhysRevD.56.3381;%%
  %332 citations counted in INSPIRE as of 08 Apr 2018
  
%\cite{Wang:2015eaa}
\bibitem{Wang:2015eaa} 
  C.~L.~Wang and R.~P.~Woodard,
  %``One-loop quantum electrodynamic correction to the gravitational potentials on de Sitter spacetime,''
  Phys.\ Rev.\ D {\bf 92}, 084008 (2015)
  doi:10.1103/PhysRevD.92.084008
  [arXiv:1508.01564 [gr-qc]].
  %%CITATION = doi:10.1103/PhysRevD.92.084008;%%
  %12 citations counted in INSPIRE as of 15 Apr 2018

%\cite{Frob:2016fcr}
\bibitem{Frob:2016fcr} 
  M.~B.~Fr\"ob and E.~Verdaguer,
  %``Quantum corrections to the gravitational potentials of a point source due to conformal fields in de Sitter,''
  JCAP {\bf 1603}, no. 03, 015 (2016)
  doi:10.1088/1475-7516/2016/03/015
  [arXiv:1601.03561 [hep-th]].
  %%CITATION = doi:10.1088/1475-7516/2016/03/015;%%
  %7 citations counted in INSPIRE as of 15 Apr 2018
  
%\cite{Radkowski:1970}
\bibitem{Radkowski:1970}
  A.~F.~Radkowski,
  %``Some aspects of the source description of gravitation,''
  Annals Phys.\  {\bf 56}, 319 (1970). 
  
%\cite{Capper:1974ed}
\bibitem{Capper:1974ed} 
  D.~M.~Capper, M.~J.~Duff and L.~Halpern,
  %``Photon corrections to the graviton propagator,''
  Phys.\ Rev.\ D {\bf 10}, 461 (1974).
  doi:10.1103/PhysRevD.10.461
  %%CITATION = doi:10.1103/PhysRevD.10.461;%%
  %103 citations counted in INSPIRE as of 16 Apr 2018 
            
%\cite{Tsamis:1992xa}
\bibitem{Tsamis:1992xa} 
  N.~C.~Tsamis and R.~P.~Woodard,
  %``The Structure of perturbative quantum gravity on a De Sitter background,''
  Commun.\ Math.\ Phys.\  {\bf 162}, 217 (1994).
  doi:10.1007/BF02102015
  %%CITATION = doi:10.1007/BF02102015;%%
  %114 citations counted in INSPIRE as of 16 Apr 2018
  
%\cite{Woodard:2004ut}
\bibitem{Woodard:2004ut} 
  R.~P.~Woodard,
  %``de Sitter breaking in field theory,''
  gr-qc/0408002.
  %%CITATION = GR-QC/0408002;%%
  %52 citations counted in INSPIRE as of 16 Apr 2018
  
%\cite{Leonard:2013xsa}
\bibitem{Leonard:2013xsa} 
  K.~E.~Leonard and R.~P.~Woodard,
  %``Graviton Corrections to Vacuum Polarization during Inflation,''
  Class.\ Quant.\ Grav.\  {\bf 31}, 015010 (2014)
  doi:10.1088/0264-9381/31/1/015010
  [arXiv:1304.7265 [gr-qc]].
  %%CITATION = doi:10.1088/0264-9381/31/1/015010;%%
  %25 citations counted in INSPIRE as of 16 Apr 2018
  
%\cite{Glavan:2013jca}
\bibitem{Glavan:2013jca} 
  D.~Glavan, S.~P.~Miao, T.~Prokopec and R.~P.~Woodard,
  %``Electrodynamic Effects of Inflationary Gravitons,''
  Class.\ Quant.\ Grav.\  {\bf 31}, 175002 (2014)
  doi:10.1088/0264-9381/31/17/175002
  [arXiv:1308.3453 [gr-qc]].
  %%CITATION = doi:10.1088/0264-9381/31/17/175002;%%
  %18 citations counted in INSPIRE as of 16 Apr 2018

%\cite{Leonard:2012fs}
\bibitem{Leonard:2012fs} 
  K.~E.~Leonard and R.~P.~Woodard,
  %``Graviton Corrections to Maxwell's Equations,''
  Phys.\ Rev.\ D {\bf 85}, 104048 (2012)
  doi:10.1103/PhysRevD.85.104048
  [arXiv:1202.5800 [gr-qc]].
  %%CITATION = doi:10.1103/PhysRevD.85.104048;%%
  %21 citations counted in INSPIRE as of 16 Apr 2018  
  
%\cite{Wang:2014tza}
\bibitem{Wang:2014tza} 
  C.~L.~Wang and R.~P.~Woodard,
  %``Excitation of Photons by Inflationary Gravitons,''
  Phys.\ Rev.\ D {\bf 91}, no. 12, 124054 (2015)
  doi:10.1103/PhysRevD.91.124054
  [arXiv:1408.1448 [gr-qc]].
  %%CITATION = doi:10.1103/PhysRevD.91.124054;%%
  %13 citations counted in INSPIRE as of 16 Apr 2018
  
%\cite{Starobinsky:1986fx}
\bibitem{Starobinsky:1986fx} 
  A.~A.~Starobinsky,
  %``Stochastic De Sitter (inflationary) Stage In The Early Universe,''
  Lect.\ Notes Phys.\  {\bf 246}, 107 (1986).
  doi:10.1007/3-540-16452-9\_6
  %%CITATION = doi:10.1007/3-540-16452-9_6;%%
  %131 citations counted in INSPIRE as of 16 Apr 2018

%\cite{Woodard:2005cw}
\bibitem{Woodard:2005cw} 
  R.~P.~Woodard,
  %``A Leading logarithm approximation for inflationary quantum field theory,''
  Nucl.\ Phys.\ Proc.\ Suppl.\  {\bf 148}, 108 (2005)
  doi:10.1016/j.nuclphysbps.2005.04.056
  [astro-ph/0502556].
  %%CITATION = doi:10.1016/j.nuclphysbps.2005.04.056;%%
  %69 citations counted in INSPIRE as of 16 Apr 2018

%\cite{Tsamis:2005hd}
\bibitem{Tsamis:2005hd} 
  N.~C.~Tsamis and R.~P.~Woodard,
  %``Stochastic quantum gravitational inflation,''
  Nucl.\ Phys.\ B {\bf 724}, 295 (2005)
  doi:10.1016/j.nuclphysb.2005.06.031
  [gr-qc/0505115].
  %%CITATION = doi:10.1016/j.nuclphysb.2005.06.031;%%
  %129 citations counted in INSPIRE as of 16 Apr 2018

%\cite{Starobinsky:1994bd}
\bibitem{Starobinsky:1994bd} 
  A.~A.~Starobinsky and J.~Yokoyama,
  %``Equilibrium state of a selfinteracting scalar field in the De Sitter background,''
  Phys.\ Rev.\ D {\bf 50}, 6357 (1994)
  doi:10.1103/PhysRevD.50.6357
  [astro-ph/9407016].
  %%CITATION = doi:10.1103/PhysRevD.50.6357;%%
  %359 citations counted in INSPIRE as of 16 Apr 2018
  
%\cite{Miao:2006pn}
\bibitem{Miao:2006pn} 
  S.~P.~Miao and R.~P.~Woodard,
  %``Leading log solution for inflationary Yukawa,''
  Phys.\ Rev.\ D {\bf 74}, 044019 (2006)
  doi:10.1103/PhysRevD.74.044019
  [gr-qc/0602110].
  %%CITATION = doi:10.1103/PhysRevD.74.044019;%%
  %110 citations counted in INSPIRE as of 16 Apr 2018
    
%\cite{Prokopec:2007ak}
\bibitem{Prokopec:2007ak} 
  T.~Prokopec, N.~C.~Tsamis and R.~P.~Woodard,
  %``Stochastic Inflationary Scalar Electrodynamics,''
  Annals Phys.\  {\bf 323}, 1324 (2008)
  doi:10.1016/j.aop.2007.08.008
  [arXiv:0707.0847 [gr-qc]].
  %%CITATION = doi:10.1016/j.aop.2007.08.008;%%
  %115 citations counted in INSPIRE as of 16 Apr 2018
  
%\cite{Miao:2008sp}
\bibitem{Miao:2008sp} 
  S.~P.~Miao and R.~P.~Woodard,
  %``A Simple Operator Check of the Effective Fermion Mode Function during Inflation,''
  Class.\ Quant.\ Grav.\  {\bf 25}, 145009 (2008)
  doi:10.1088/0264-9381/25/14/145009
  [arXiv:0803.2377 [gr-qc]].
  %%CITATION = doi:10.1088/0264-9381/25/14/145009;%%
  %57 citations counted in INSPIRE as of 17 May 2018
  
%\cite{Kitamoto:2010et}
\bibitem{Kitamoto:2010et} 
  H.~Kitamoto and Y.~Kitazawa,
  %``Non-linear sigma model in de Sitter space,''
  Phys.\ Rev.\ D {\bf 83}, 104043 (2011)
  doi:10.1103/PhysRevD.83.104043
  [arXiv:1012.5930 [hep-th]].
  %%CITATION = doi:10.1103/PhysRevD.83.104043;%%
  %31 citations counted in INSPIRE as of 17 May 2018
  
%\cite{Kitamoto:2011yx}
\bibitem{Kitamoto:2011yx} 
  H.~Kitamoto and Y.~Kitazawa,
  %``Infra-red effects of Non-linear sigma model in de Sitter space,''
  Phys.\ Rev.\ D {\bf 85}, 044062 (2012)
  doi:10.1103/PhysRevD.85.044062
  [arXiv:1109.4892 [hep-th]].
  %%CITATION = doi:10.1103/PhysRevD.85.044062;%%
  %31 citations counted in INSPIRE as of 17 May 2018
        
%\cite{Miao:2005am}
\bibitem{Miao:2005am} 
  S.~P.~Miao and R.~P.~Woodard,
  %``The Fermion self-energy during inflation,''
  Class.\ Quant.\ Grav.\  {\bf 23}, 1721 (2006)
  doi:10.1088/0264-9381/23/5/016
  [gr-qc/0511140].
  %%CITATION = doi:10.1088/0264-9381/23/5/016;%%
  %95 citations counted in INSPIRE as of 17 May 2018
  
%\cite{Miao:2006gj}
\bibitem{Miao:2006gj} 
  S.~P.~Miao and R.~P.~Woodard,
  %``Gravitons Enhance Fermions during Inflation,''
  Phys.\ Rev.\ D {\bf 74}, 024021 (2006)
  doi:10.1103/PhysRevD.74.024021
  [gr-qc/0603135].
  %%CITATION = doi:10.1103/PhysRevD.74.024021;%%
  %63 citations counted in INSPIRE as of 17 May 2018
  
%\cite{Glavan:2015ura}
\bibitem{Glavan:2015ura} 
  D.~Glavan, S.~P.~Miao, T.~Prokopec and R.~P.~Woodard,
  %``Graviton Loop Corrections to Vacuum Polarization in de Sitter in a General Covariant Gauge,''
  Class.\ Quant.\ Grav.\  {\bf 32}, no. 19, 195014 (2015)
  doi:10.1088/0264-9381/32/19/195014
  [arXiv:1504.00894 [gr-qc]].
  %%CITATION = doi:10.1088/0264-9381/32/19/195014;%%
  %9 citations counted in INSPIRE as of 17 May 2018    

%\cite{Prokopec:2002uw}
\bibitem{Prokopec:2002uw} 
  T.~Prokopec, O.~Tornkvist and R.~P.~Woodard,
  %``One loop vacuum polarization in a locally de Sitter background,''
  Annals Phys.\  {\bf 303}, 251 (2003)
  doi:10.1016/S0003-4916(03)00004-6
  [gr-qc/0205130].
  %%CITATION = doi:10.1016/S0003-4916(03)00004-6;%%
  %114 citations counted in INSPIRE as of 08 Apr 2018
  
%\cite{Leonard:2012si}
\bibitem{Leonard:2012si} 
  K.~E.~Leonard, T.~Prokopec and R.~P.~Woodard,
  %``Covariant Vacuum Polarizations on de Sitter Background,''
  Phys.\ Rev.\ D {\bf 87}, no. 4, 044030 (2013)
  doi:10.1103/PhysRevD.87.044030
  [arXiv:1210.6968 [gr-qc]].
  %%CITATION = doi:10.1103/PhysRevD.87.044030;%%
  %13 citations counted in INSPIRE as of 08 Apr 2018
  
%\cite{Leonard:2012ex}
\bibitem{Leonard:2012ex} 
  K.~E.~Leonard, T.~Prokopec and R.~P.~Woodard,
  %``Representing the Vacuum Polarization on de Sitter,''
  J.\ Math.\ Phys.\  {\bf 54}, 032301 (2013)
  doi:10.1063/1.4793987
  [arXiv:1211.1342 [gr-qc]].
  %%CITATION = doi:10.1063/1.4793987;%%
  %13 citations counted in INSPIRE as of 08 Apr 2018 

\end{thebibliography}
\end{document}